\documentclass[twocolumn,twoside,slac_two]{revtex4}
\usepackage{graphicx}
\usepackage{fancyhdr}
\usepackage{hyperref}
\hypersetup{
    colorlinks=true,
    urlcolor=magenta,
}
\begin{document}

%Title of paper
\title{Spatial distribution of high-energy protons in the inner radiation belt on the data of low Earth orbit space experiments}

% Repeat the \author .. \affiliation  etc. as needed
%
% \affiliation command applies to all authors since the last
% \affiliation command. The \affiliation command should follow the
% other information

\author{S.Yu. Aleksandrin, A.M. Galper, S.V. Koldashov, M.A.  Mayorova, T.R. Zharaspayev}
\affiliation{National Research Nuclear University MEPhI (Moscow Engineering Physics Institute)}

\begin{abstract}
Measurements of the ARINA instrument on board the Resurs-DK1 satellite (altitude $\sim$ 600 km and inclination $\sim$ 70$^{o}$, since 2006 till 2016) and the VSPLESK instrument on board the International Space Station (altitude $\sim$ 400 km and inclination $\sim$ 52$^{o}$, since 2008 till 2013) in low Earth orbits were presented in this report. Both instruments are identical in terms of physical layout. They can measure high-energy protons in the range 30-100 MeV with 10\% energy resolution and 7$^{o}$ angular accuracy. Data analysis was carried out for the total period of proton flux measurement by the instruments. L-B proton distributions in the inner radiation belt (L$<$2) were studied in dependence on proton energy. Geographical and pitch-angle distribution of proton intensity were studied for chosen L-shells. These distributions were analyzed during the decreasing part of the 23$^{rd}$ solar cycle and the main part of the 24$^{th}$ one.
\end{abstract}

%\maketitle must follow title, authors, abstract
\maketitle

\thispagestyle{fancy}

% body of paper here - Use proper section commands
% References should be done using the \cite, \ref, and \label commands
% Put \label in argument of \section for cross-referencing
%\section{\label{}}

\section{Introduction}
The radiation belts have been known practically since the first spacecraft launches \cite{bib:bibl1, bib:bibl2}. They have become a great threat to the operation of instruments located on the satellites. When the low Earth orbit satellites flying in the South Atlantic Anomaly, they cross the inner radiation belt (RB). The radiation load increases sharply in thousand times in this area. The inner radiation belt consists mostly of protons with energies of several tens of MeV. The RB proton fluxes may vary in time. There are various causes influencing on particle fluxes in the RB, as external (solar activity) and internal (geomagnetic field). The study of fluxes is both scientific and practical interest.

Direct measurements of proton fluxes began almost from the first flight and continue regular up to now. However for this time the basic data are receiving from the GOES satellite series which are in geostationary orbit \cite{bib:bibl3}. In this paper we consider the data obtained during two experiments ARINA and VSPLESK which were a long time in low-Earth orbits.

\section{Instruments}

ARINA and VSPLESK spectrometers have the same physical layout \cite{bib:bibl4}. The main part of the spectrometer is a multi-layered scintillation detector. The instrument measures fluxes of electrons with energies of 3-30 MeV and protons with energies of 30-100 MeV. To determine the energy consider only those particles that have stopped in the detector, i.e. the bottom counter is included as anticoincidence. Defining the layer where the particle is stopped, it is possible to estimate the energy. spectrometer resolution for electrons 15\%, and for protons - about 10\%. The angular resolution is $\pm$ 7$^{o}$.

Spectrometer ARINA was set on  Resurs-DK1 spacecraft, which had been launched in summer 2006 on orbit with an inclination $\sim$70$^{o}$ and altitude 350-600 km. In autumn 2009, the spacecraft's orbit was changed to a circular with altitude $\sim$ 600 km. The satellite operated until January 2016.

VSPLESK was set on ISS (an inclination $\sim$52$^{o}$ and altitude  $\sim$400 km) in the autumn of 2008 and operated until the end of 2013.

%\begin{figure}
%\includegraphics[width=65mm]{JACpic_mc.eps}
%\caption{Layout of papers.}
%\label{l2ea4-f1}
%\end{figure}

\section{Fluxes of protons in the SAA}
Data fluxes of protons in the South Atlantic anomaly are essential for determining the radiation dose on the low Earth orbit satellites and instruments on them.

\begin{figure}
\includegraphics[width=85mm]{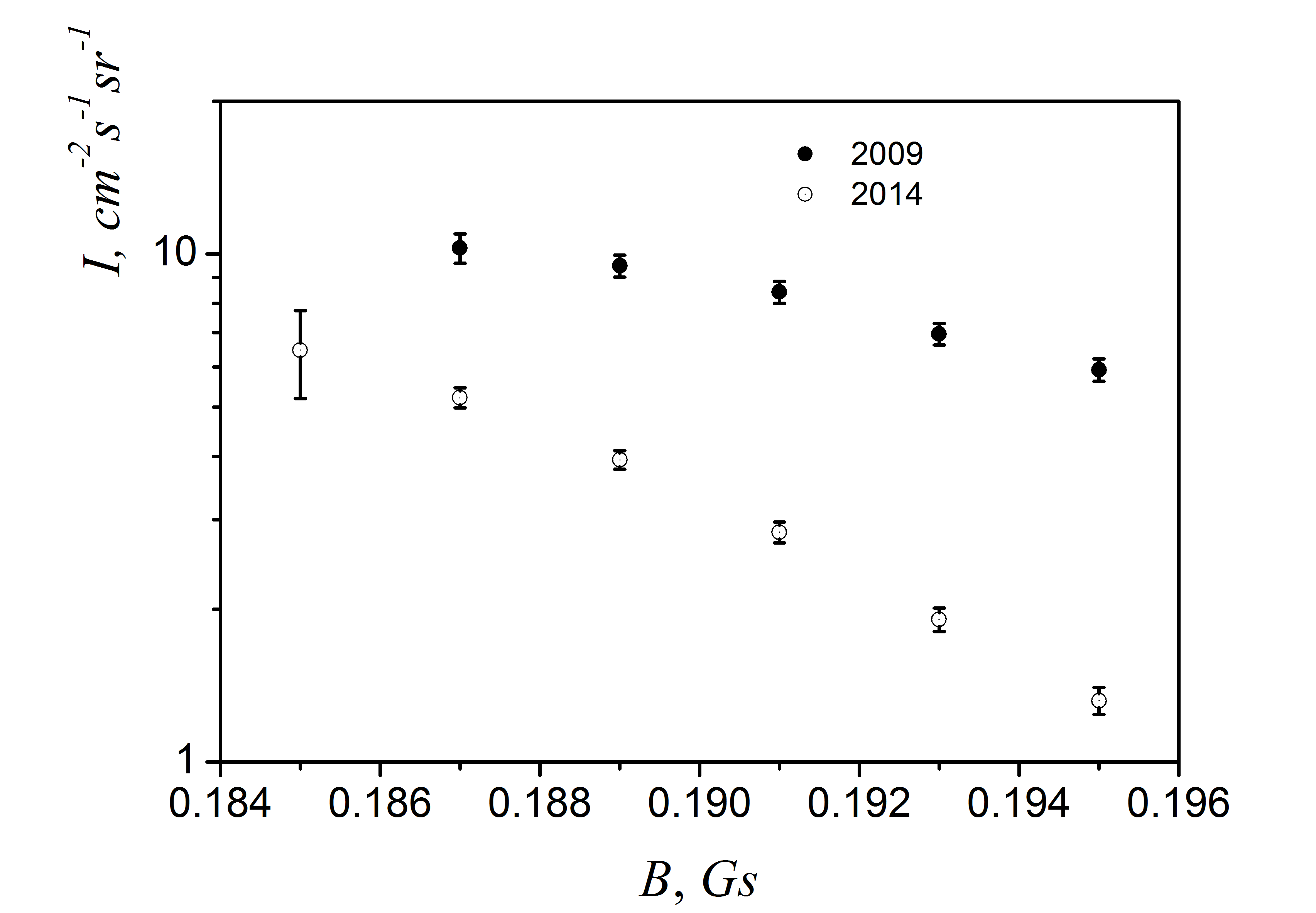}
\caption{Proton flux (E=45-55MeV) for L=1.16 (ARINA 2006-2016).}
\label{fig1}
\end{figure}

These fluxes may have a short-term variations and long-term changes. In this paper we studied the fluxes of protons with energies from 30 to 100 MeV, which were registered at altitudes from 400 (spectrometer VSPLESK) up to 600 km (spectrometer ARINA). Depending on fluxes were obtained from the magnetic field of the Earth for a variety of L-shells from 1.16 to 2. Figure~\ref{fig1} shows the corresponding dependence on the proton energy $\sim$ 50 MeV for L=1.16. Data are presented for 2009 and 2014, which are periods of maximum and minimum solar activity, respectively (minimal flux in the period of solar  maximum).

There has been a decrease in flux by increasing the intensity of the geomagnetic field, i.e. reduction of the height of the mirror points of the trapped particles, and, respectively, and a decrease in the time of their life while longitude drift around the Earth. For observed in experiments L-shells, the value of the geomagnetic induction B is relatively large for and AP8 model \cite{bib:bibl5} gives great error for flux estimation, while maintaining the trend - decrease of flux with increasing geomagnetic induction along the shell.

\begin{figure}
\includegraphics[width=85mm]{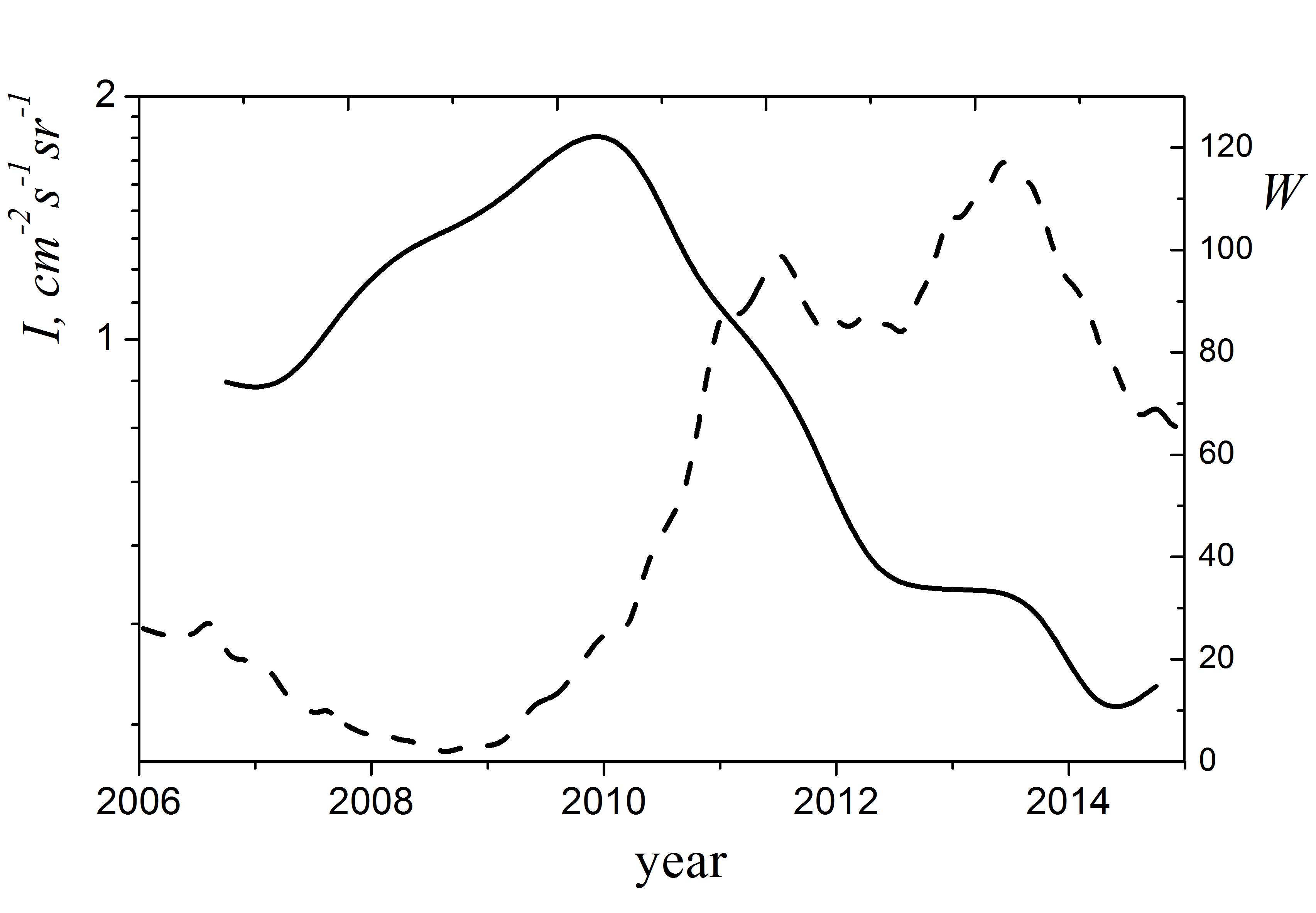}
\caption{Proton flux time variation (L=1.16, E=45-55MeV, ARINA instrument). The dashed curve is the Wolf number.}
\label{fig2}
\end{figure}

Figure~\ref{fig2} shows the proton flux during years of observation (solid line) for selected L-shell in a specified interval of the magnetic field of the Earth along the shell. For L=1.16 maximum proton  flux in the solar maximum is several times less than the minimum period of solar activity. The same figure shows the number of sunspots, characterized Wolf number (dashed line).

It is evident that during the phase of maximum solar activity, characterized by the maximum number of Wolf, fluxes of trapped particles decreases in 5-7 times relatively the quiet Sun period in the phase of solar activity minimum. These changes were observed by ARINA spectrometer. VSPLESK instrument for the same L-shell observed flux variation only 2 times, as the ISS, on board of which was set VSPLESK, had an orbit lower than Resurs-DK1 satellite (ARINA), that is geomagnetic field intensity was higher, and the particle mirror point fall below. With increasing energy protons detected difference between the periods of solar maximum and minimum decreases.

% figures should be put into the text as floats.
% Use the graphics or graphicx packages (distributed with LaTeX2e)
% and the \includegraphics macro defined in those packages.
% See the LaTeX Graphics Companion by Michel Goosens, Sebastian Rahtz,
% and Frank Mittelbach for instance.
%
% Here is an example of the general form of a figure:
% Fill in the caption in the braces of the \caption{} command. Put the label
% that you will use with \ref{} command in the braces of the \label{} command.
% Use the figure* environment if the figure should span across the
% entire page. There is no need to do explicit centering.

% \begin{figure}
% \includegraphics{}%
% \caption{\label{}}
% \end{figure}

% Surround figure environment with turnpage environment for landscape
% figure
% \begin{turnpage}
% \begin{figure}
% \includegraphics{}%
% \caption{\label{}}
% \end{figure}
% \end{turnpage}

%\begin{figure*}[t]
%\centering
%\includegraphics[width=135mm]{fig1.png}
%\caption{Proton flux (E=45-55MeV) for L=1.16 (ARINA 2006-2016) } \label{ecrs_2016_logo}
%\end{figure*}

\section{Conclusion}

The results of the processing data from two satellite experiments ARINA and VSPLESK were analyzed. The instruments had the same physical layout, but they were onboard on different spacecrafts. Analysis showed the dependence of proton flux intensity of solar activity. The proton flux which is registered by ARINA spectrometer can increase depending on L-shell up to 5-7 times (L = 1.16) during the Solar minimum. At the same time the proton flux in VSPLESK experiment increased only 2 times on the same L-shell, because VSPLESK was installed on the ISS, whose orbit is lower than Resurs-DK1 satellite (ARINA) one.

% If in two-column mode, this environment will change to single-column
% format so that long equations can be displayed. Use
% sparingly.
%\begin{widetext}
% put long equation here
%\end{widetext}

% If you have acknowledgments, this puts in the proper section head.
\bigskip % extra skip inserted
\begin{acknowledgments}
The authors wish to thank JACoW for their guidance in preparing
this template.

This work was supported by National Research Nuclear University MEPhI in the framework of the Russian Academic Excellence Project (contract No. 02.a03.21.0005, 27.08.2013).
\end{acknowledgments}

\bigskip % extra skip inserted
% Create the reference section using BibTeX:
%\bibliography{basename of .bib file}

\end{document}